\documentclass{ws-ijmpe}

\begin{document}

\markboth{Jason Glyndwr Ulery For the STAR Collaboration}{ARE THERE MACH CONES IN HEAVY ION COLLISIONS?  THREE-PARTICLE CORRELATIONS FROM STAR}

\catchline{}{}{}{}{}

\title{ARE THERE MACH CONES IN HEAVY ION COLLISIONS?\\
THREE-PARTICLE CORRELATIONS FROM STAR}

\author{\footnotesize JASON GLYNDWR ULERY FOR THE STAR COLLABORATION}

\address{Department of Physics, Purdue University, 525 Northwestern Avenue\\
West Lafayette, Indiana 47907, USA
\\ulery@physics.purdue.edu}

\maketitle

\begin{history}
\received{(received date)}
\revised{(revised date)}
\end{history}

\begin{abstract}

We present results from STAR on 3-particle azimuthal correlations for a $3<p_T<4$ GeV/c trigger particle with two softer $1<p_T<2$ GeV/c particles.  Results are shown for pp, d+Au and high statistics Au+Au collisions at $\sqrt{s_{NN}}=200 GeV$.  We observe a 3-particle correlation in central Au+Au collisions which may indicate the presence of conical emission.  In addition, the dependence of the observed signal angular position on the $p_T$ of the associated particles can be used to distinguish conical flow from simple QCD-\v{C}erenkov radiation.  An important aspect of the analysis is the subtraction of combinatorial backgrounds. Systematic uncertainties due to this subtraction and the flow harmonics $v_2$ and $v_4$ are investigated in detail.

\end{abstract}

\section{Introduction}
Heavy ion collisions create a medium that may be the quark gluon plasma (QGP).  This medium can be studied through jets and jet-correlations.  Jets make a good probe because their properties can be calculated in the vacuum with perturbative quantum chromodynamics (pQCD).  Two-particle jet-like azimuthal correlations have shown the away-side shape in central Au+Au collisions to be broadened with respect to $pp$ and peripheral Au+Au collisions or even double humped \cite{star2p,phenix2p} (see Fig.~\ref{fig:Fig1}a).  The away-side structure is consistent with many different physics mechanisms including: large angle gluon radiation \cite{gluon,gluon2}, jets deflected by radial flow or preferential selection of particles due to path-length dependent energy loss, hydrodynamic conical flow generated by Mach-cone shock waves \cite{mach1,mach2}, and \v{C}erenkov radiation \cite{cerenkov,cerenkov2}.  Three-particle correlations can be used to differentiate conical flow and \v{C}erenkov radiation, which have the characteristic of conical emission, from other mechanisms.  In addition, the associated particle $p_T$ dependence of the conical emission angle can be used to differentiate between hydrodynamic conical flow and simple \v{C}erenkov radiation. 

\section{Analysis Procedure}
The 3-particle correlation analysis method is rigorously described in \cite{proceed}.  The results reported here are for charged trigger particles of $3<p_T<4$ GeV/c and two charged associated particles of $1<p_T<2$ GeV/c (except where otherwise noted).  The data were all taken in the STAR time projection chamber for {\it pp}, d+Au and Au+Au collisions at $\sqrt{s_{NN}}$=200 GeV/c.  

\begin{figure}[htbp]
	\centering
		\includegraphics[width=1.0\textwidth]{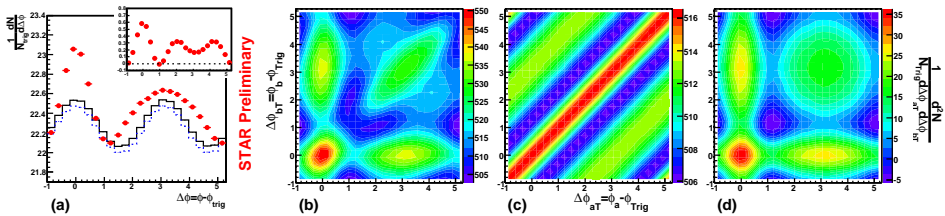}
			\vspace*{-0.4cm}
	\caption{(color online) (a) Raw 2-particle correlation (points), background from mixed events with flow modulation added-in (solid) and scaled by ZYA1 (dashed), and background subtracted 2-particle correlation (insert).  (b) Raw 3-particle correlation, (c) soft-soft background, $\beta \alpha^{2}B_{3}^{inc}$ and (d) hard-soft background + trigger flow, $\hat{J}_{2}\otimes \alpha B_{2}^{inc}$ + $\beta \alpha^{2}B_{3}^{inc,TF}$.  See text for detail.  Plots are from ZDC-triggered 0-12\% Au+Au collisions at $\sqrt{s_{NN}}$=200 GeV/c.}
	\label{fig:Fig1}
\end{figure}

Figure~\ref{fig:Fig1}b shows the raw 3-particle azimuthal distribution in $\Delta\phi_{aT}=\phi_{a}-\phi_{T}$ and $\Delta\phi_{bT}=\phi_{b}-\phi_{T}$ where $\phi_{T}$, $\phi_a$, and $\phi_b$ are the azimuthal angles of the trigger particle and the two associated particles respectively. Combinatorial backgrounds must be removed to obtain the genuine 3-particle correlation signal.  The analysis is performed by treating the events as composed of two components, particles that are jet-like correlated with the trigger particle and background particles.
One source of background, the hard-soft background, results when one of the associated particles has a jet-like correlation with the trigger particle and the other is uncorrelated, except for the correlation due to flow.  The background is constructed from the 2-particle jet-like correlation, $\hat{J}_{2}$, folded with the normalized 2-particle background, $B_{2}^{inc}$, Fig.~\ref{fig:Fig1}a.  The 2-particle background is constructed by mixing events with the flow modulation added in pairwise from the average $v_2$ values from the measurements based on the reaction plane and 4-particle cumulant methods \cite{star2p}.  For the $v_4$ contribution we use the parameterization $v_{4}=1.15v_{2}^{2}$ from the data\cite{reactionplane}.  The background is normalized (with scale factor $\alpha$) to the signal within $0.8<|\Delta\phi|<1.2$ (zero yield at 1 radian or ZYA1).  We shall refer to the hard-soft background as $\hat{J}_{2}\otimes \alpha B_{2}^{inc}$. 

Another source of background, the soft-soft background, results from correlations between the two associated particles  which are independent of the trigger particle.  This background is obtained from mixed events, where the trigger particle and the associated particles are from different events in the same centrality window.  We shall refer to the soft-soft background as $B_{3}^{inc}$.  Since the two associated particles are from the same event, this background contains all of the correlations between the two associated particles that are independent of the trigger particle, including correlations from minijets, other jets in the event, and flow.  

The flow between the two associated particles that is independent of the trigger particle was accounted for in the soft-soft term, but particles are also correlated with the trigger particle via flow.  The trigger flow is added in triplet-wise from mixed events, where the trigger and associated particles are all from different events in the same centrality window.  The $v_{2}$ and $v_{4}$ values are obtained the same way as for the 2-particle background.  The number of triplets is determined from the inclusive events.   We shall refer to the backgrounds from trigger flow as $B_{3}^{inc,TF}$.  
The total background is then, $\hat{J}_{2}\otimes \alpha B_{2}^{inc}$ + $\beta \alpha^{2}(B_{3}^{inc}+B_{3}^{inc,TF})$.  Both $B_{3}^{inc}$ and $B_{3}^{inc,tf}$ are scaled by $\beta \alpha^2$.  The normalization $\alpha^2$ corrects for the multiplicity bias from requiring a trigger particle.  The factor $\beta$ accounts for the effect of non-poission multiplicity distributions and is obtained such that the number of associated pairs in the background subtracted jet-like three-particle correlation signal equals the square of the number of associated particles in the background subtracted jet-like two-particle correlation signal. 
Figure~\ref{fig:Fig1}c and d show $\beta \alpha^{2}B_{3}^{inc}$ and $\hat{J}_{2}\otimes \alpha B_{2}^{inc}$ + $\beta \alpha^{2}B_{3}^{inc,TF}$, respectively.  

\section{Results}

\begin{figure}[htbp]
	\centering
		\includegraphics[width=.95\textwidth]{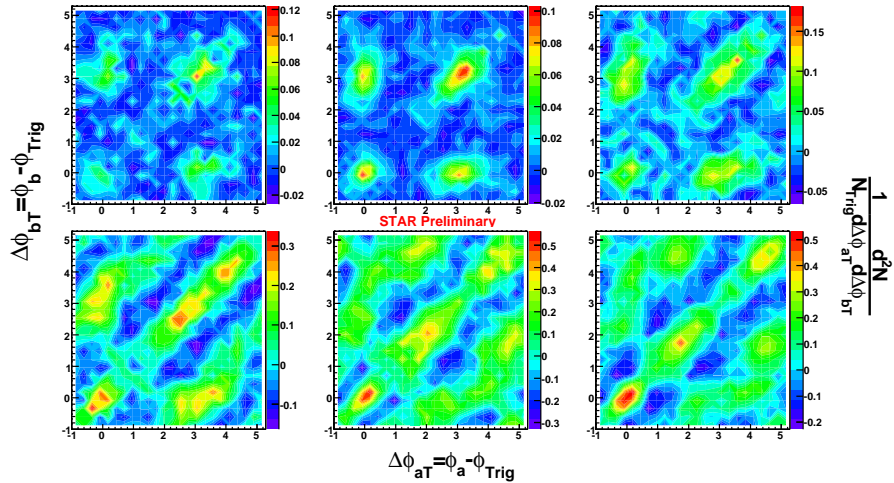}
			\vspace*{-0.3cm}
	\caption{(color online) Background subtracted 3-particle correlations for {\it pp} (top left), d+Au (top middle), and Au+Au 50-80{\%} (top right), 30-50{\%} (bottom left), 10-30{\%} (bottom center), and ZDC triggered 0-12{\%} (bottom right) collisions at $\sqrt{s_{NN}}$=200 GeV/c.}
	\label{fig:Fig3}
\end{figure}

Figure~\ref{fig:Fig3} shows background subtracted 3-particle jet-like correlation signals.  The {\it pp} and d+Au results are similar.  Peaks are clearly visible for the near-side, (0,0), the away-side, ($\pi$,$\pi$) and the two cases of one particle on the near-side and the other on the away-side, (0,$\pi$) and ($\pi$,0). The away-side peak displays diagonal elongation that is consistent with $k_T$ broadening.  The perpherial Au+Au results show additional on-diagonal elongation of the away-side peak which may be due to contribution from deflected jets.  The additional on-diagonal broadening persists into the more central Au+Au collisions.  In addition, the more central Au+Au collisions display an off-diagonal structure, at about $\pi\pm1.45$ radians, that is consistent with conical emission.  The structure increases in magnitude with centrality and is prominent in the high statistics top 12\% central data provided by the on-line ZDC trigger.

\begin{figure}[htbp]
\hfill
\begin{minipage}[t]{.49\textwidth}
	\centering
		\includegraphics[width=1.0\textwidth]{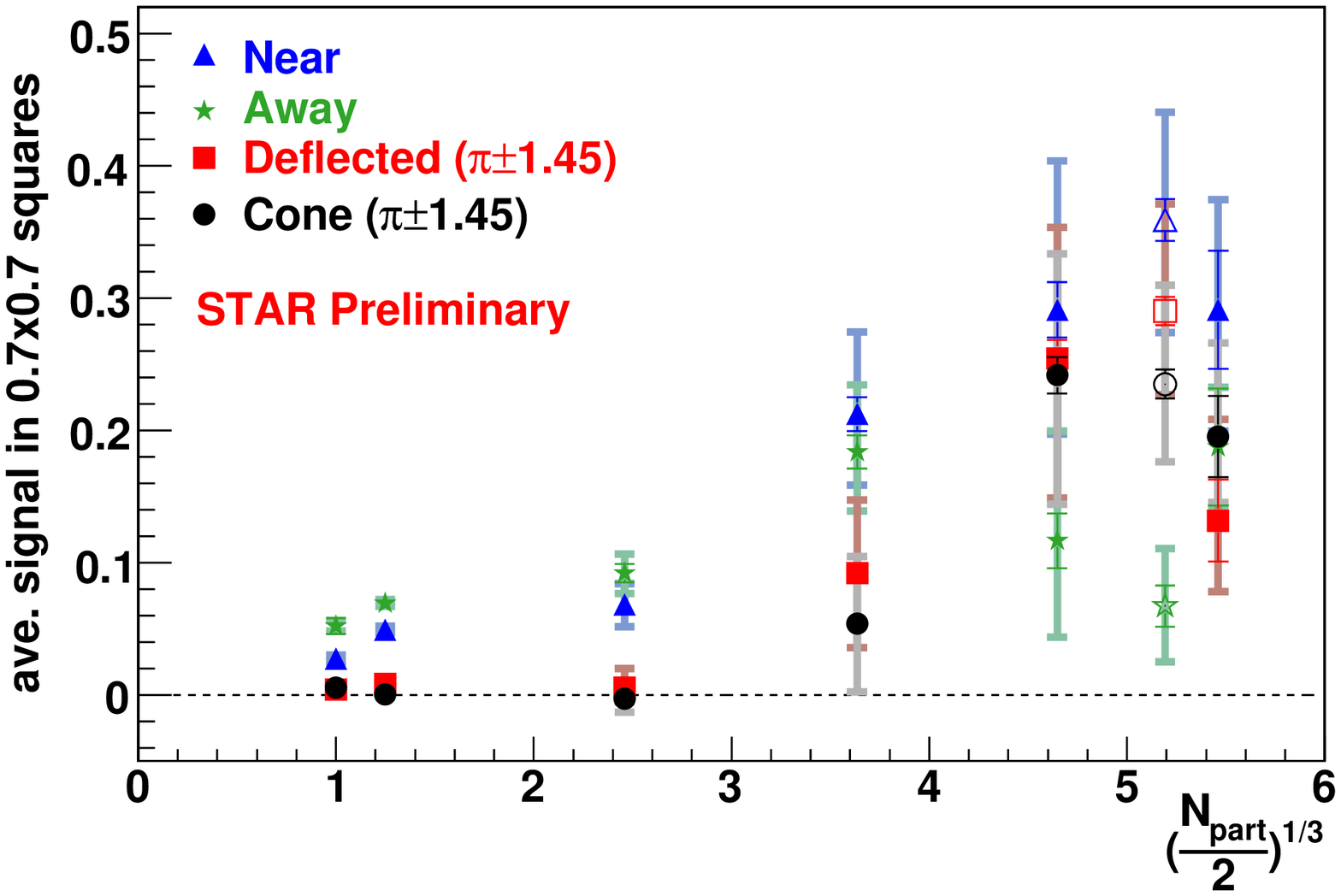}
			\end{minipage}
	\hfill
\begin{minipage}[t]{.49\textwidth}
	\includegraphics[width=1.0\textwidth]{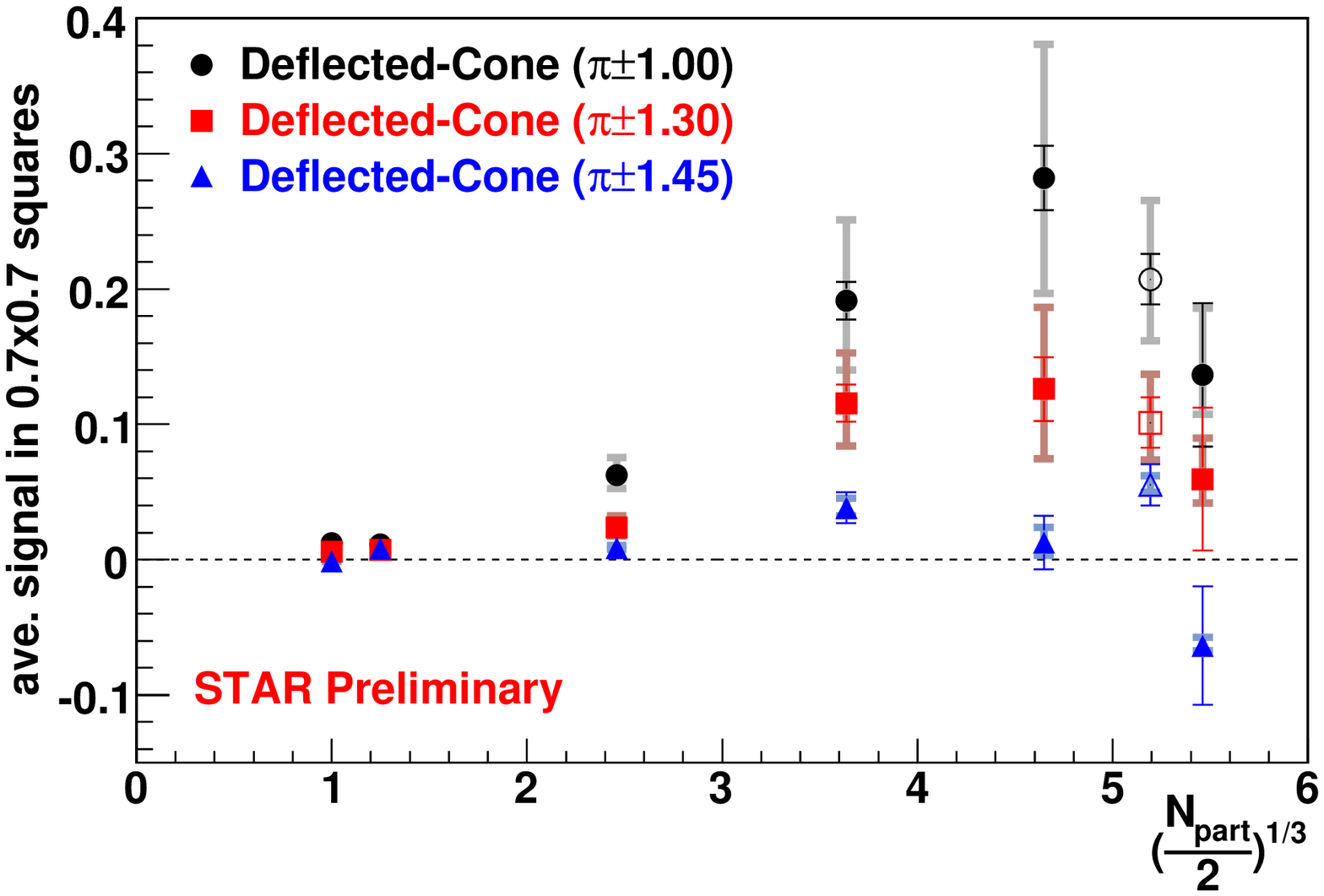}
			\end{minipage}
	\hfill
			\vspace*{-0.3cm}
	\caption{(color online) (left) Average signals in 0.7 $\times$ 0.7 boxes at (0,0) (triangle), ($\pi$,$\pi$) (star), ($\pi\pm1.45$,$\pi\pm1.45$) (square), and ($\pi\pm1.45$,$\pi\mp1.45$) (circle).  (right) Differences in average signals, between ($\pi\pm1.45$,$\pi\pm1.45$) and ($\pi\pm1.45$,$\pi\mp1.45$) (triangle), between ($\pi\pm1.3$,$\pi\pm1.3$) and ($\pi\pm1.3$,$\pi\mp1.3$) (square), and between ($\pi\pm1.0$,$\pi\pm1.0$) and ($\pi\pm1.0$,$\pi\mp1.0$) (circle).  Solid error bars are statistical and shaded are systematic.  $N_{part}$ is the number of participants.  The ZDC 0-12\% points (open symbols) are shifted to the left for clarity.}
	\label{fig:Fig4}	
\end{figure}

\begin{figure}[htbp]
	\centering
		\includegraphics[width=.90\textwidth]{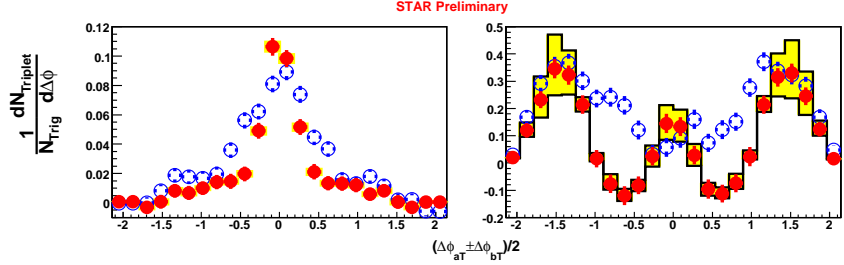}
			\vspace*{-0.35cm}
	\caption{(color online) Away-side projections of a strip of width 0.7 radians for (left) d+Au and (right) 0-12\% ZDC Triggered Au+Au.  Off-diagonal projection (solid) is $(\Delta\phi_1-\Delta\phi)/2$ and on-diagonal projection (open) is $(\Delta\phi_1+\Delta\phi)/2-\pi$. Shaded bands are systematic errors.}
	\label{fig:Fig5}
\end{figure}

Figure~\ref{fig:Fig4} (left) shows the centrality dependence of the average signal strengths in different regions.  The off-diagonal signals (circle) increase with centrality and significantly deviate from zero in central Au+Au collisions.  The locations of the off-diagonal signals were determined from a double Gaussian fit to a strip projected to the off-diagonal, Fig.~\ref{fig:Fig5}, and were found to be 1.45 radians from $\pi$.  
The differences between on-diagonal signals, where both conical emission and deflected jets may contribute, and off-diagonal signals, where only conical emission contributes is shown in figure~\ref{fig:Fig4} (right).  Since conical emission signals are expected to be of equal magnitude on-diagonal as off-diagonal, the difference may indicate the contribution from deflected jets.  The difference decreases with distance from ($\pi$,$\pi$).

\begin{figure}[htbp]
\hfill
\begin{minipage}[t]{.47\textwidth}
	\centering
		\includegraphics[width=1\textwidth]{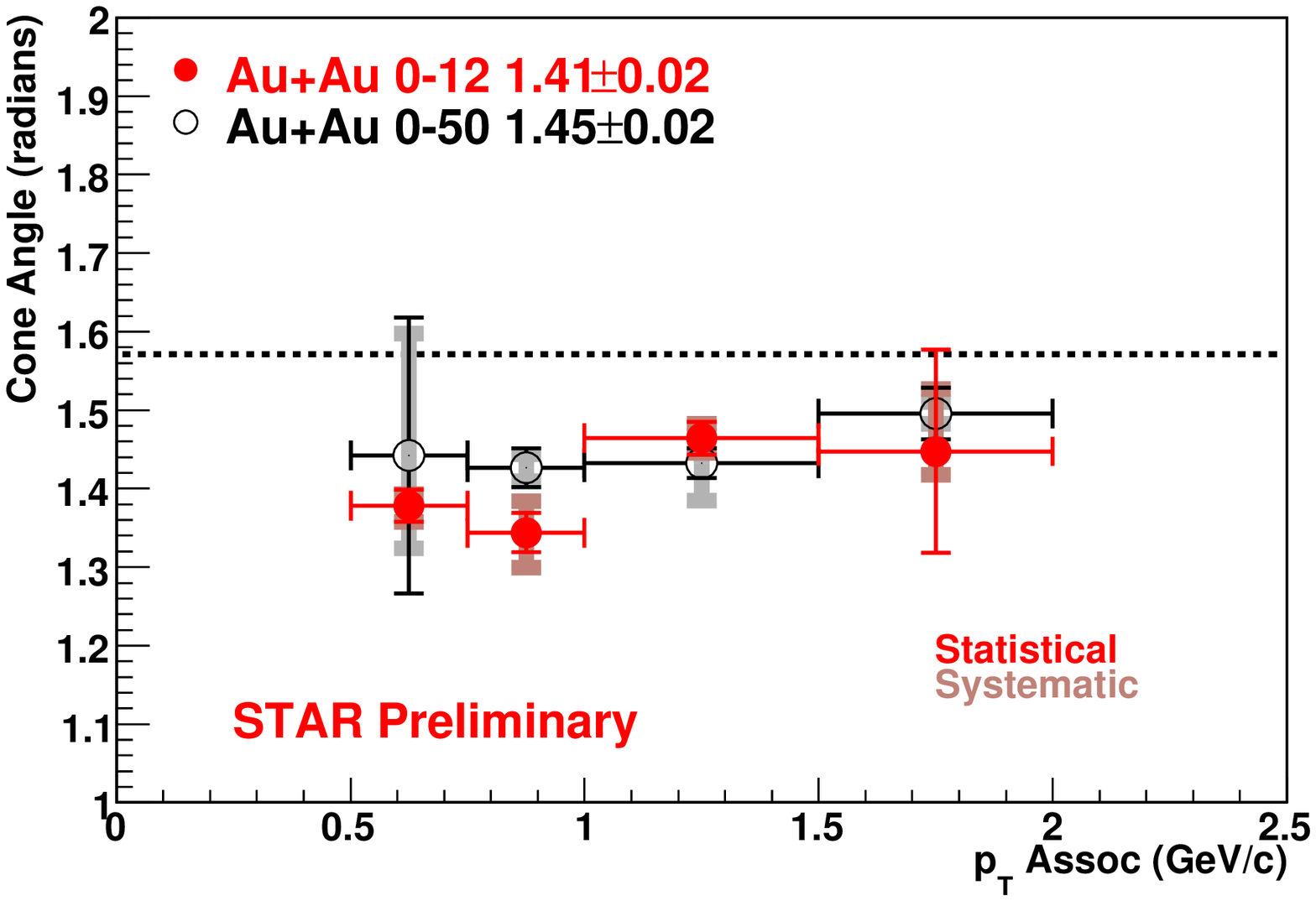}
			\end{minipage}
	\hfill
\begin{minipage}[t]{.47\textwidth}
	\includegraphics[width=1\textwidth]{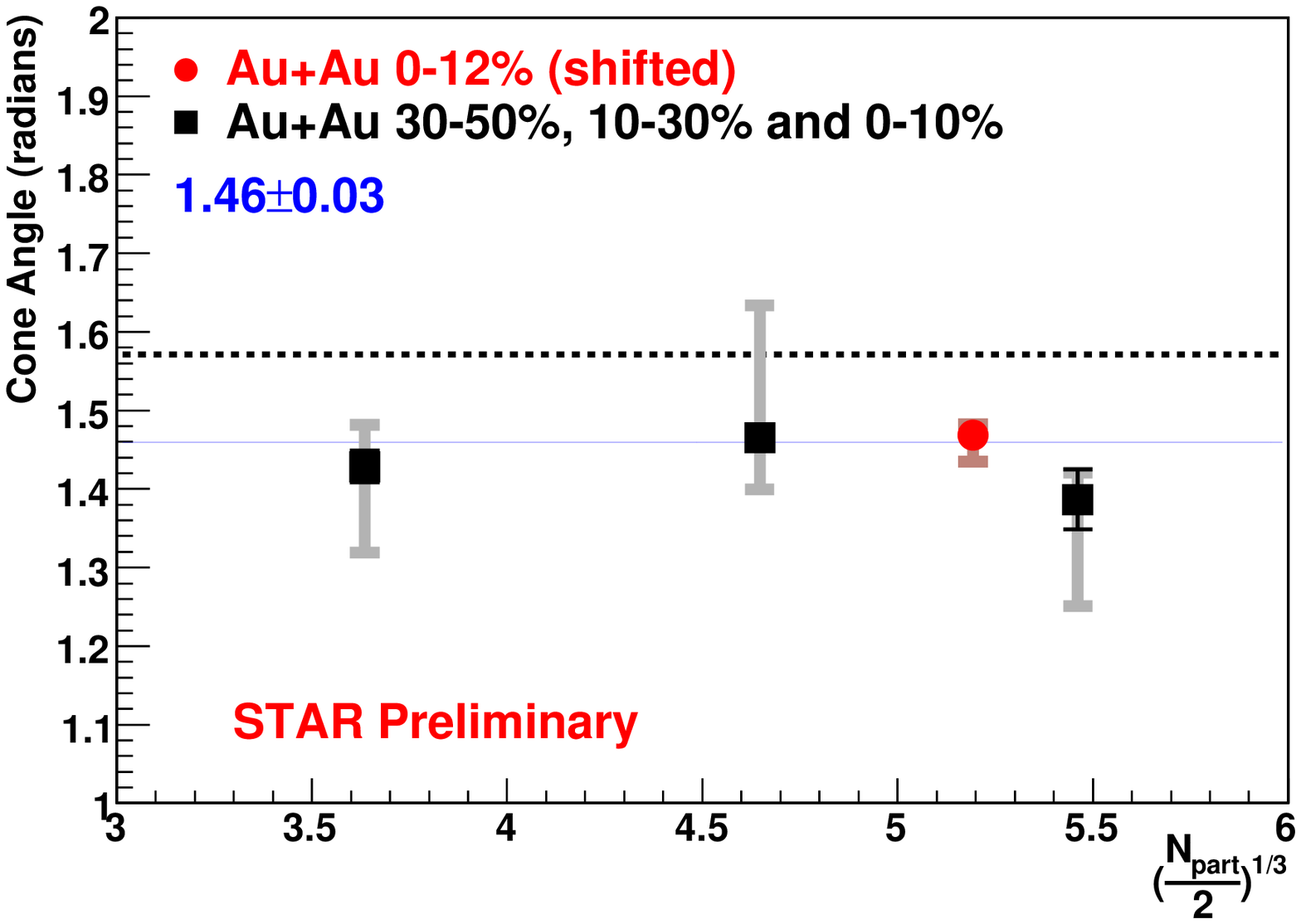}
			\end{minipage}
	\hfill
			\vspace*{-0.4cm}
	\caption{(color online) Emission angles from double Gaussian fits.  (left) Angle as a function of associated particle $p_T$ for Au+Au 0-12\% ZDC triggered data (filled) and Au+Au 0-50\% from minimum bias data (open). (right) Angle as a function of centrality for Au+Au 0-12\% ZDC triggered data (circle) and Au+Au 30-50\%, 10-30\% and 0-10\% from minimum bias data (square).  The 0-12\% point has been shifted for clarity.  Corresponding off-diagonal peak values from fits to a constant are indicated in the legends.  The dashed line is at $\pi/2$.  Solid error bars are statistical and shaded are systematic.}
	\label{fig:Fig6}	
\end{figure}

The Mach cone emission angle is expected to be independent of the associated particle momentum \cite{mach2}, while the \v{C}erenkov radiation model in Ref.~\cite{cerenkov} predicts an emission angle that is sharply decreasing with increasing associated particle momentum.  Figure~\ref{fig:Fig6} (left) shows the dependence the off-diagonal peak angle on associated particle $p_T$.  The angle is consistent with constant as a function of associated particle $p_T$.  

Figure~\ref{fig:Fig6} (right) shows the centrality dependence the off-diagonal peak angle.  The angle is consistent with remaining constant as a function of centrality for mid-central and central Au+Au collisions.  The solid line at 1.46 on the plot is from a fit to a constant.  

\section{Systematics}
\begin{figure}[htbp]
	\centering
		\includegraphics[width=.9\textwidth]{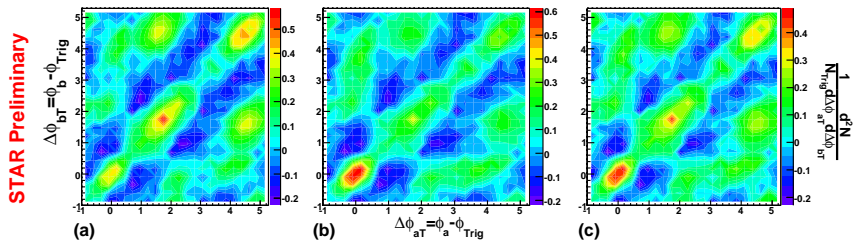}
			\vspace*{-0.3cm}
	\caption{(color online) 0-12\% Au+Au ZDC triggerd data for different systematic checks: (a) reaction plane $v_2$, (b) 4-particle cumulant $v_2$, and (c) normalization region for $\alpha$ of $0.6<|\Delta \phi|<1.4$}
	\label{fig:Fig7}
\end{figure}

The major sources of systematic error are the elliptic flow measurement and the normalization.  The default $v_2$ used is the average those measured by the reaction plane and 4-particle cumulant methods.  We use the reaction plane and 4-particle cumulant $v_2$ as the upper and lower bounds to estimate the systematic uncertainty of the $v_2$ subtraction.  Figure~\ref{fig:Fig7}a and b show the background subtracted 3-particle correlation for reaction plane and 4-particle $v_2$ respectively.  The signal is robust with respect to this variation.  The hard-soft background and trigger flow backgrounds individually vary a great deal with the change in elliptic flow but the variations cancel to first order in the sum.  

To study the effect of the normalization the size of the normalization window was doubled to $0.6<|\Delta\phi|<1.4$.  The signal is robust with respect to this change in normalization.  
Other sources of systematic error include the effect on the trigger particle flow from requiring a correlated particle (±20\% on trigger particle $v_2$), uncertainty in the $v_4$ parameterization, and multiplicity bias effects on the soft-soft background.  The systematic errors shown in figures include all sources mentioned.

\section{Conclusion}
Three-particle azimuthal correlations have been studied for trigger particles of $3<p_T<4$ GeV/c and associated particles of $1<p_T<2$ GeV/c in {\it pp}, d+Au, and Au+Au collisions at $\sqrt{s_{NN}}$=200 GeV/c by STAR.   This analysis treats events as the sum of two components, particles that are jet-like correlated with the trigger and background particles.  On-diagonal broadening has been observed in {\it pp} and d+Au consistent with $k_T$ broadening.  Additional on-diagonal broadening has been seen in Au+Au collisions possibly due to deflected jets.  Off-diagonal peaks consistent with conical emission are present in  central Au+Au collisions.  To discriminate between Mach cone emission and simple \v{C}erenkov gluon radiation, a study of the associated particle $p_T$ dependence was performed.  No strong $p_T$ dependence of the angles on associated particle $p_T$ is observed.  This result is consistent with Mach cone emission.

\end{document}